# Servo Actuating System Control Using Optimal Fuzzy Approach Based on Particle Swarm Optimization

Dev Patel, Li Jun Heng, Abesh Rahman, Deepika Bharti Singh

*Abstract*—This paper presents a new optimal fuzzy approach based on particle swarm optimization (PSO) evolutionary algorithm for controlling the servo actuating system. It is clear that attaining the maximum stability margin is the prominent goal in control design of servo actuating systems. To reach the control goal, two main steps of design are required; an appropriate identification method and a controller development. Hence, the nonlinear system is first identified by the fuzzy algorithm. Then, the controller parameters and the algorithm's weighting functions are tuned through the Particle Swarm Optimization algorithm. The objective function of optimal control strategy is such that the minimum error between the actual and the identified data is attained. The effectiveness of the proposed approach comparing to the conventional fuzzy control with regular parameter tuning is illustrated and analyzed in the simulations.

## I. Introduction

The servo actuators are widely useful and applicable in industries, control, automation area, and manufacturing processes [1]. Considering that these systems have nonlinear dynamics and parameter variations, it is difficult to design a proper controller for them. Various control methods have been used on these systems in the previous researches like real time iterative learning control, adaptive control, neural network control, etc [2-4].

Regular certain mathematical models have been used for representing the servo systems' nonlinear models. They have been used based on the nonlinear control theory. The basic method in these algorithms is to linearize the exact system through the nonlinear state feedback. Although using the certain mathematical models, some important factors such as payload variations and mass flow parameters' uncertainties are being neglected in the model. To overcome this issue, fuzzy approaches are being widely used for servo actuators' control.

Extensive research works have been done in the identification area of servo systems in the past. Actually the nonlinear servo systems are assumed to be presented by local models through fuzzy algorithms. One of the famous fuzzy methods used before is the Takagi & Sugeno fuzzy model, in which smoothed piecewise linear models are used to analyze and synthesize the nonlinear system. In fact, fuzzy controllers are very simple to implement on servo systems, also they build an efficient nonlinear fuzzy controller for extensive range of applications [5], [6], [7]. Fuzzy approaches and neuro-fuzzy control methods are known to be very efficient for applying on servo actuators and various researchers applied fuzzy algorithms to design the controller or to identify the nonlinear system based on fuzzy methods.

In implementing the fuzzy algorithm, choosing appropriate weights is of great importance. Several evolutionary algorithms have been used in order to choose the best parameters for the fuzzy structure [8]. The objective in the optimal algorithm is to minimize the error between the actual and identified model such that high performance and robustness of the system are achieved properly.

In order to implement the fuzzy functions and fuzzy controllers, an appropriate programming environment is needed to be chosen. The fuzzy toolbox of MATLAB has been chosen in this work since it is convenient, simple to be used and very detailed.

This paper work is focused on identifying the nonlinear servo actuator system in the first stage through the fuzzy structure. In the second stage, he parameters of the fuzzy structure and the weighting functions of the algorithm are then tuned and set based on the Particle Swarm Optimization (PSO) algorithm. However Genetic Algorithm (GA) is very famous in tuning the controller's parameters, PSO is proved to operate faster in such applications since its calculations are simpler than GA. As a matter of fact, PSO is one of the most efficient techniques for adapting parameters of fixed order controllers [9]. The literature in [9] is effectively used in this work as the work have applied PSO on H infinity loop shaping controller design for the mechanical beam system.

Thus, by PSO optimization algorithm the optimum parameters and the weighting functions for designing the fuzzy control design are attained. The aim is to minimize the objective function including the errors between the actual and identified models. The simulations results proved the efficiency and the effectiveness of the proposed algorithm.

The paper is organized as follows. The system dynamics and its details are represented in section II. Section III describes the chosen fuzzy control design approach. Section IV gives the Particle Swarm Optimization (PSO) algorithm description used in the proposed technique. Then, the results of the approach on the servo actuating system and its comparison to the conventional method results are shown in section V, and the last section gives the conclusion as well.

## II. Dynamic Modeling of Servo Actuating System

The system under consideration is a servo actuator or a servo drive as Fig. 1.

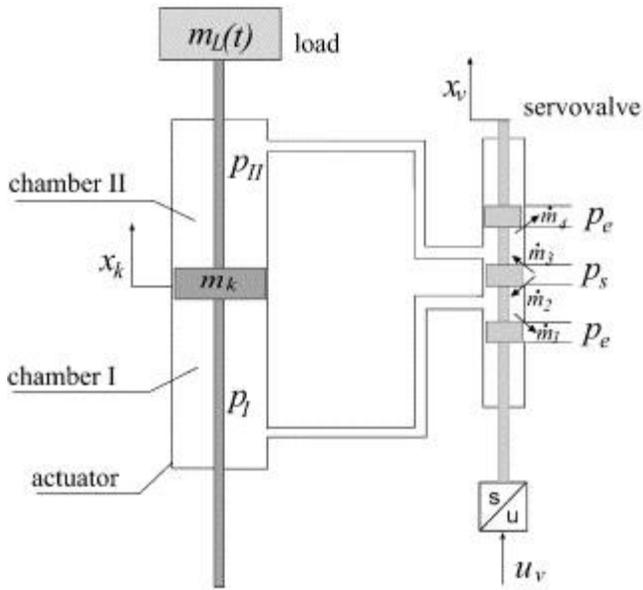

Fig. 1: Servo actuating system

The main components of the system are the cylinder, valve, load, actuator, also the fuzzy controller is to be added to the structure in the next part. The valve controls the flow rate of the compressed air. The position of the cylinder is being controlled by the valve. The position of the valve is used to control the spool movement, which controls the air flow rate of the cylinder and also controls the velocity and position of the cylinder.

The mathematical model of the servo actuating system is attained by physics laws. The descriptive model of the system a nonlinear model of order five. The advantage of this model is that it considers the mass flow rate, the pressure dynamics, the friction, the motion dynamics. Hence, the state space equations are stated as (1).

$$\dot{x} = v$$

$$\dot{v} = (\frac{1}{M+m})(A_1 P_1 - A_2 P_2 - F_{friction} + mg\,sin\theta)$$

$$\dot{P_1} = (\varepsilon/A(l+x))(Cf(P_1, u) - A_1 P_1 v - \Delta h)$$

$$\dot{P_2} = (\varepsilon/A(l-x))(-Cf(P_2, u) - A_2 P_2 v - \Delta h) \quad (1)$$

where $v$ is the velocity, $x$ is the position, and $P_1$ and $P_2$ are the absolute pressure of cylinder.
$M$ and $m$ are the piston and load mass respectively. Also, $A_1$ and $A_2$ are the areas of the chambers. $l$ is the length of cylinder, $F_{friction}$ is the force from friction, g is the gravity acceleration, $\varepsilon$ is the coefficient, and $C$ is the constant.

### III. Fuzzy Control Design

The controller's model chosen for this system is an adaptive fuzzy controller from literature [5]. To implement this controller, first the part of the system that can be linearized is linearized. The linearized equation for the system would be as (2).

$$\dot{x}_1 = x_2$$
$$\dot{x}_2 = x_3$$
$$\vdots$$
$$\dot{x}_n = f(x_1, \cdots, x_n) + g(x_1, \cdots, x_{n2})u(t) + d(t)$$

$$y = x_2 \quad (2)$$

Note that the functions $f$ and $g$ are unknown functions. Using the control signal, the system parameters $\theta$ are tuned. The feedback controller is designed by Mamdani fuzzy system as [5].

By tuning the parameters vector $\theta$, the unknown functions $f$ and $g$ are estimated as $\hat{f}$ and $\hat{g}$. The objective is to track the desired output trajectory in order to minimize the tracking error.

If-then fuzzy rules are used to estimate the functions $f$ and $g$. The fuzzy rules are designed based on the input-output behavior of the system. The parameter vector $\theta$ is also split into two different vectors $\theta_f$ and $\theta_g$ for the functions estimation through the algorithm.

Thus the control signal can be stated as (3).

$$u = \left(\frac{1}{\hat{g}(x,\theta)}\right)[-\hat{f}(x,\theta_f) + y^n + K^T e] \quad (3)$$

Note that the error vector $e$ in the above equation is the difference between the system output and the desired output trajectory.

Therefore, the following stages are introduced to implement fuzzy control strategy on the system:

- Step 1: The input output behavior of the system is considered by the $P$ number of the input fuzzy sets and the $q$ number of output fuzzy sets.
- Step 2: The if then statements are built base on the system's input output behavior.

Worth mentioning that the level of if then fuzzy rules accuracy is related to the human knowledge of the system. The if then statements are as follows:
if $x_i$ is $A_i$, then $\hat{f}(x,\theta_f)$ is a member of $B_i$.

For estimating $f$ and $g$ functions, singleton fuzzifier is used and the adaptive law for estimating fuzzy functions would be as:

$$\hat{f}(x,\theta_f) = \theta_f^T \varepsilon_f(X)$$

$$\hat{g}(x,\theta_f) = \theta_f^T \varepsilon_f(X) \quad (4)$$

Consider that the type of fuzzy approach is the indirect adaptive fuzzy method.

The membership functions are picked from the MATLAB fuzzy toolbox as Fig. 2 (the triangular-shaped membership functions (trimf of fuzzy toolbox) are used for the inputs and the outputs of the system).

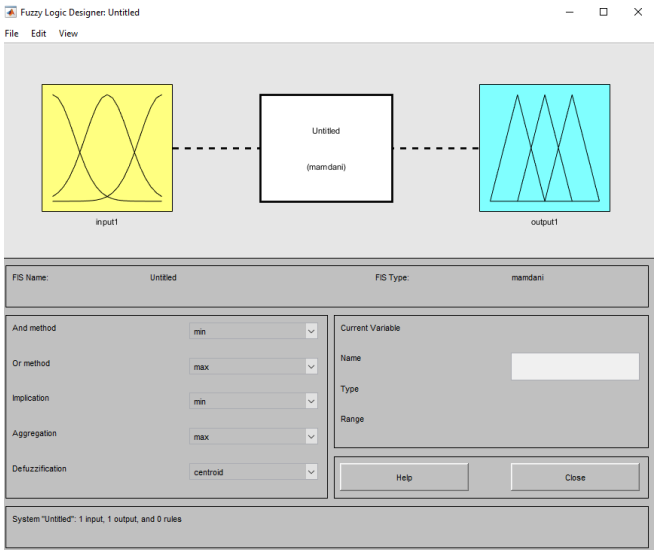

Fig. 2: Membership functions and Mamdani inference engine in Fuzzy toolbox

Also, the membership functions (MF1 to MF10) for one of the inputs is shown as Fig. 3.

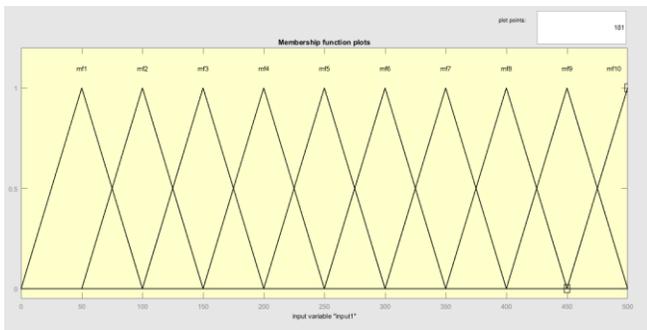

Fig. 3: Membership functions for the fuzzy system input

The output membership function (MF0 to MF13) is also supposed as Fig. 4.

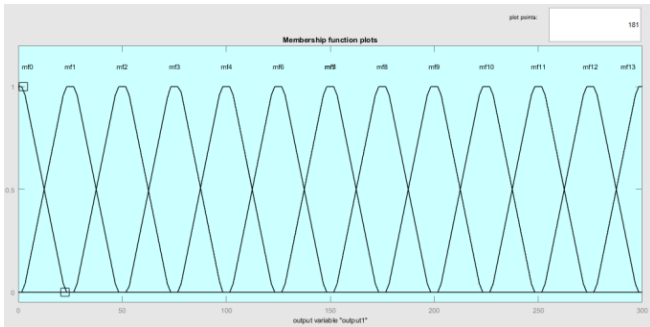

Fig. 4: Membership functions for the fuzzy system output

The if then fuzzy rule created are also from the following window in Fig. 5.

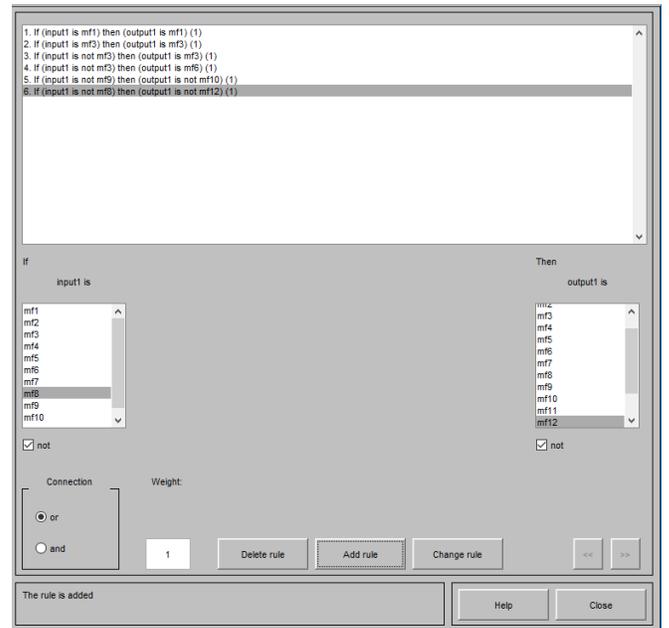

Fig. 5: if-then fuzzy rules window in the toolbox

Choosing the P matrix as $10^3 \times I$ also the block diagram of the fuzzy structure is as follows [5]:

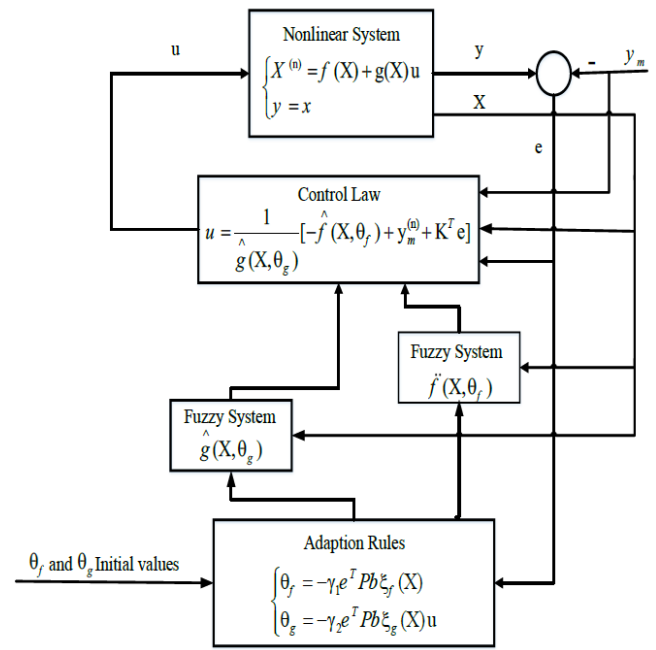

Fig. 6: Fuzzy control structure

Using the algorithm and the fuzzy strategy the controller's main structure is defined. In the next step, the structure's parameters are optimized based on the PSO optimization algorithm (Particle Swarm Optimization).

The closed loop diagram of the fuzzy controller and the system is illustrated in Fig. 7.

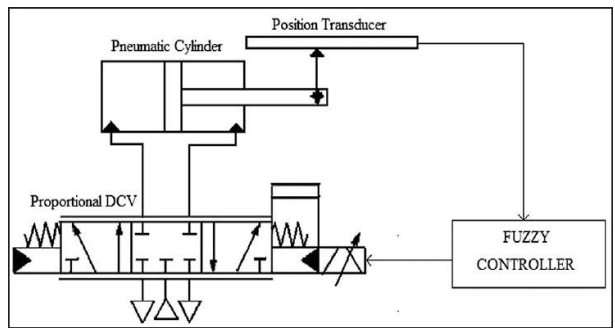

Fig. 7: The closed loop system of the servo actuating system and the fuzzy controller

## IV. PSO OPTIMIZATION ALGORITHM

Particle Swarm Optimization (PSO) is a generalized optimization technique based on the insects' social interactions [10]. The following diagram shows step by step of the PSO algorithm.

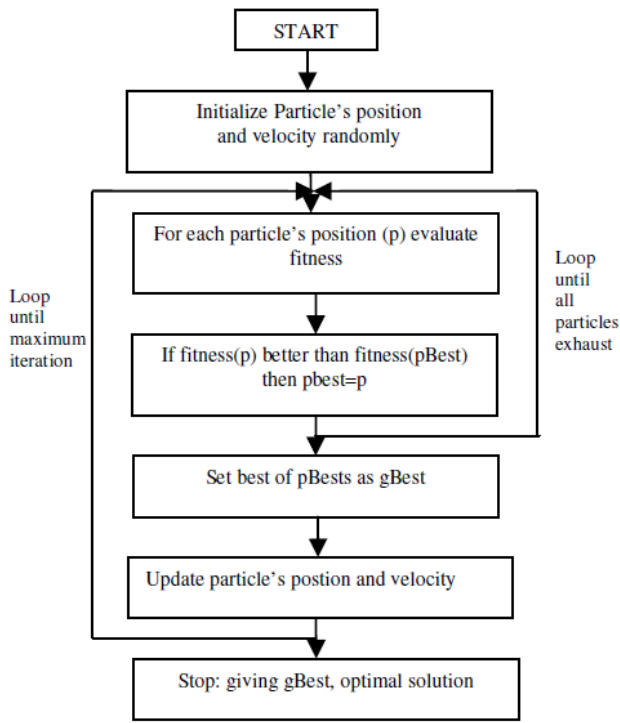

Fig. 8: Particle Swarm Optimization algorithm flowchart

In PSO, each particle has its own position and the velocity of the particles are being updated based on an algorithm that determines the direction and velocity of the movement of each particle. Each position participates in evaluating the fitness (objective) function. After each iteration, new candidate particles are created with the aim to attain the optimum point in the fitness function. The velocity and the position of each particle is updated based on equations (5) and (6) respectively.

$$v_{k+1}^i = w v_k^i + c_1 rand(p^i - x_k^i) + c_2 rand(p_k^g - x_k^i) \quad (5)$$

$$x_{k+1}^i = x_k^i + v_{k+1}^i \quad (6)$$

in the above equations, $v_k^i$ is the particle's velocity at iteration $k$ and the $v_{k+1}^i$ is the particle's velocity at iteration $k+1$. The other parameters are the inertia weight, particle and swarm constant coefficients, particle's individual best and global best.

In fact, for each optimization algorithm like PSO there exists a stopping criterion to recognize where to stop the iterations; the iterations stop where the convergence is attained in the strategy.

In fact, the main use of PSO in fuzzy control design is for choosing the scaling gains, the controller structure settings, determining the membership functions of the inputs and outputs, fuzzifier and defuzzifire, etc. Therefore, PSO is used mainly to find the optimum fuzzy controller for the specific system.

It is worth mentioning that PSO is similar to Genetic algorithm (GA) in nature; since these two algorithms are population base, however GA is more time consuming compared to PSO while used for nonlinear problems. This is the main reason here PSO (not GA) is recommended in designing fuzzy model.

## V. SIMULATION RESULTS

In this section, the simulation results of using the proposed fuzzy method based on PSO are represented. Also for evaluating the results, the conventional fuzzy control design method without optimization process is used on the system.

Fig. 9 shows the results using the classic fuzzy control on servo actuating system and Fig. 10 shows the results of implementing the proposed method.

Comparing Fig. 9 and Fig. 10, the velocity tracking error using the proposed PSO fuzzy controller is recognizably lower than the tracking error using the conventional fuzzy controller. As it can be seen from Fig. 9, the trajectory starts from 10 and ends with less than 15, however the desired trajectory starts from 9 and ends in more than 15. Regarding Fig. 10, the estimated trajectory perfectly follows the desired trajectory.

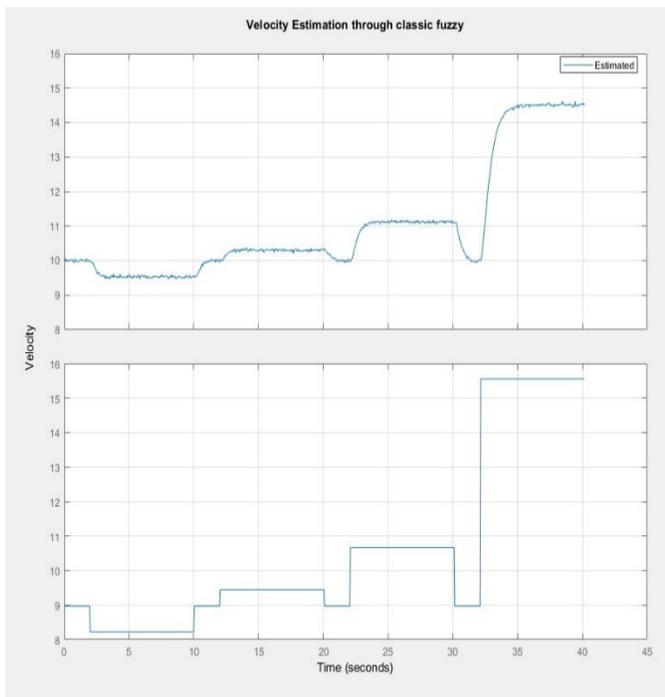

Fig. 9: System velocity trajectory regarding the desired trajectory using classic fuzzy control

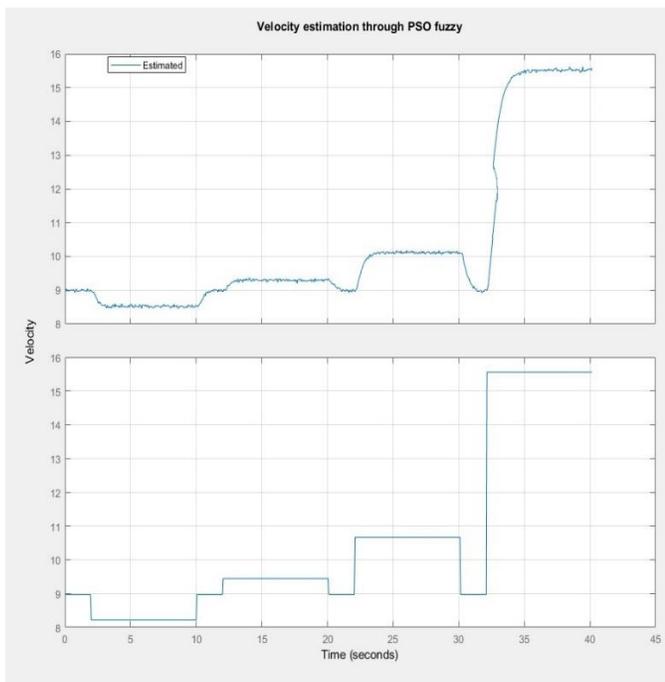

Fig. 10: System velocity trajectory regarding the desired trajectory using PSO fuzzy control

## VI. CONCLUSIONS

This paper proposed a new a new fuzzy approach based on Particle Swarm Optimization algorithm for control design of a servo actuating system. Since the system descriptive model is nonlinear, fuzzy identification is used as the controller and the model structure. The parameters of the controller are then tuned base on PSO optimization strategy. PSO is chosen as the optimization approach because it is computationally efficient and simple for nonlinear problems. The objective of the optimization algorithm is such that the minimum tracking between the actual model output and the estimated model output is attained. The simulation results proved the effectiveness and efficiency of this hybrid approach in the servo system control design. Moreover, to strongly emphasize the superiority of the new method, the outcomes of the conventional fuzzy algorithm are illustrated and compared to the results from the proposed approach.

REFERENCES

[1] B. W. Anderson, *The analysis and design of pneumatic systems*, New York, London, Sydney: Wiley, 1967.
[2] Xu, J., Panda, Sanjib Kumar, Lee, Tong Heng, *Real-time Iterative Learning Control Design and Applications* (Advances in Industrial Control), 2009.
[3] H. Wenmei, Y. Young, and T. Yali, *Adaptive neuron control based on predictive model in pneumatic servo system*, 2002.
[4] K. Harbick, S. Sukhatme., *Speed control of a pneumatic Monopod using a neural network*, 2002.
[5] R. Eini and S. Abdelwahed, "Indirect Adaptive fuzzy Controller Design for a Rotational Inverted Pendulum," *2018 Annual American Control Conference (ACC)*, Milwaukee, WI, USA, 2018, pp. 1677-1682. doi: 10.23919/ACC.2018.8431796.
[6] J. Espinosa, J. Vandewalle & V. Wertz, *Fuzzy Logic, Identification and Predictive Control*, London: Springer, 2004.
[7] R. Jang, *MATLAB - Fuzzy Toolbox - The MathWorks, Inc. Revision: 1.12*, Date: 2000, 15.
[8] T. Gulrez, A. Hassanien, *Advances in Robotics and Virtual Reality* (Vol. 26, Intelligent Systems Reference Library). Berlin, Heidelberg: Springer Berlin Heidelberg, 2012.
[9] R. Eini, Flexible Beam Robust H-infinity Loop Shaping Controller Design Using Particle Swarm Optimization. Journal of Advances in Computer Research, 5(3), Quarterly pISSN: 2345-606x eISSN: 2345-6078, 2014.
[10] M. Couceiro, P. Ghamisi, *Fractional Order Darwinian Particle Swarm Optimization* (Springer Briefs in Applied Sciences and Technology). Cham: Springer International Publishing, 2016.